# Development of radioactive beams at ALTO:
# Part 1. Physicochemical comparison of different types of UC$_x$ targets using a multivariate statistical approach

Julien Guillot[a*], Sandrine Tusseau-Nenez[b], Brigitte Roussière[a], Nicole Barré-Boscher[a], François Brisset[c], Sylvain Denis[d]

[a] Institut de Physique Nucléaire d'Orsay CNRS/IN2P3 UMR 8608 - Université Paris Sud - Université Paris Saclay, F-91406 Orsay Cedex

[b] Laboratoire de Physique de la Matière Condensée, CNRS, Ecole Polytechnique - Université Paris Saclay, Route de Saclay, F-91128 Palaiseau

[c] Institut de Chimie Moléculaire et des Matériaux d'Orsay CNRS UMR 8182 - Université Paris Sud - Université Paris Saclay, F-91405 Orsay Cedex

[d] Institut de Chimie et des Matériaux Paris-Est CNRS UMR 7182 - Université Paris Est Créteil, F-94320 Thiais

* Corresponding author: guillotjulien@ipno.in2p3.fr

______________________________________________________________________

______________________________________________________________________

**Abstract:**

The optimization of the microstructure of the UC$_x$ target is a key point since many years in the field of ISOL method. The ultimate goal is to facilitate the release of the fission products, especially those with short half-lives. Fourteen UC$_x$ samples were synthetized from different uranium and carbon sources using three mixing protocols. All carburized samples were systematically characterized in terms of nature and proportion phases, grain and aggregate size, open and close porosity proportion and open pore size distribution. Our results were analysed using a multivariate statistical approach in order to remove any subjective bias. Strong correlations between the physicochemical characteristics of the samples as well as the impact of the synthesis process have been highlighted. In particular, using carbon nanotubes as carbon source combined with a new method of mixing is the key parameter to limit the sintering and to obtain samples with small grains and a high porosity well distributed over small pores. Moreover the microstructure obtained proved to be stable at high temperature.

______________________________________________________________________

Introduction:

In most ISOL facilities, the uranium carbide targets formed by a mixture of UC$_2$ and UC and called UC$_x$ have become the reference target material [1]. Indeed, in order to improve the kinetics of the diffusion and desorption processes, the target must withstand temperature as high as 2000 °C.

The use of refractory compound obtained by powder metallurgy provides fission fragments with a minimum diffusion path in the solid [2]. The UC$_x$ targets are in the form of sintered pellets composed of microscopic uranium carbide grains mixed with graphite [3].

Since the 1960s, the target synthesis control is one of the key points that allows a high production of radioactive nuclei and improve their release. These two requirements seem to be contradictory: since the first needs a high uranium density and the second a high porosity. UC$_x$ targets offer good stability at high temperature and good thermal conductivity [4]. However, high temperatures combined with high doses of radiation may alter the initial properties of the material. Frequently mentioned problems, which in the long term weaken the material, are the swelling [5]–[7] and the modifications of cell parameters [8]. In order to maximize their release properties, uranium carbide targets are used in ISOL facilities with excess carbon [3], typically U/C = 1/6 ratios [9].

The Oak Ridge Laboratory performed fission product yield measurements using uranium dicarbide targets with different U/C ratios. The authors observe that targets with the lowest excess of carbon are less efficient [10].

Very dense pellets of uranium monocarbide were synthesized at PNPI in Russia by hot pressing [11]–[13]. The very low release fractions of the fission products (FPs) measured at 1600 °C were attributed to the low open porosity (5%) in the material [14].

The role of porosity had been studied in $UO_2$ targets in 1977 by Turnbull and Friskney [15]. They established a relationship between the quantity of porosity and the release efficiency. Afterwards, many prototype targets have been developed with the aim increasing open porosity and producing interconnected porosity.

The first technical solution to obtain a porous target was to build composite uranium carbide targets, initially fiber-based [16]–[21] and more recently based on carbon nanotubes (CNT) [22], [23]. Within this approach, the desired porosity was achieved.

The second technical solution is to choose, for the uranium precursor, uranium oxalate hydrate ($U(C_2O_4)_2.6H_2O$) or triuranium octoxide ($U_3O_8$) instead of uranium dioxide ($UO_2$). During the heating process, their decomposition into $UO_2$ leads to degassing of not only CO but also $CO_2$, inducing open porosity in the material [24]–[26].

The last solution is to add an organic polymer during the green target preparation in order to increase degassing during the carburization [25], [27]–[30]. This technique allows producing an interconnected and architectured porosity according to the polymer selected.

In the ISOL community, some studies have been dedicated to the control of sintering conditions [17], [23], [31], [32]. In particular they have found that grain growth reduces the porosity. In order to prevent abnormal grain growth during sintering, they recommend to control the temperature and to use homogeneous and nanoscaled powder size, or to insert nanosized carbon between the grains.

In the context of nuclear fuel studies, it has been shown that in $UO_2$ the FPs produced after irradiation are localized in the grains, both at the edge and at the heart of the grains. [33]. The release of FPs by $UO_2$ and UC can be assumed to be similar since diffusion in both substances is similar [34]. The theoretical study of Forsberg and Massih demonstrated the correlation between the $UO_2$ grains size and the FPs release [35].

In powders, as in bulk materials, two mechanisms are involved: the diffusion of the FPs towards the grain surface and the effusion through the porosity. The importance of each parameter depends on the particles size and the porous structure of the sample. Greene *et al.* showed that release properties are improved when the $UC_2$ grain size (for a target with a U/C ratio of 1/8) is reduced from 250 to 44 μm [36].

Most of the above-mentioned studies involved grains with micrometric sizes. Going to a nanometric structure, where the diffusion paths will be minimized, should all the more improve the release. Some release simulations of different isotopes by $UC_x$ targets were carried out with the RIBO code in order to study the release dependence upon grain size and porosity [37], [38]. In the case of isotopes with short sticking time, reducing the size of the grains improves the FPs release. On the other hand, in the case of isotopes with medium or long sticking time, using larger grains has the advantage of limiting the sticking time. Finally, whatever the isotope sticking time, increasing the porosity has a positive impact on the FPs release.

These previous studies show that many parameters have to be optimized in order to obtain a target that provides an efficient release of the FPs, in particular of those with very short half-life. Our present study follows the preliminary tests conducted at IPNO [14], [39]. A systematic work on the manufacturing parameters (grinding and mixture of precursor powders, pressing, carburization ...) led to the development of synthesis protocols for nanostructured targets. Fourteen different samples were prepared and characterized. A multivariate statistical approach is used to establish correlations between the structural properties of the samples, namely the porosity and the pore size distribution, the $UC_x$ grain and aggregate size, and factors involved in the manufacturing process, such as the material density (and therefore the C/U ratio), the nature of the carbon and uranium precursors and the mixing and synthesis protocols.

1. Synthesis of the samples

1.1. Precursor powders

Two types of uranium powders were used: the first, a uranium oxide powder (AREVA, sample No. 65496) containing 0.25 wt% of $^{235}U$, with impurities of Cr, Ni and Fe respectively 4, 6 and 16 µg/gU. This powder was dry milled for 4h following the protocol described in Guillot *et al*. [40]. The second one, a uranium oxalate hydrate powder was prepared at IPNO [41], [42] and milled with the same conditions as the $UO_2$ one but only for 2 h.

We used three different carbon sources: graphite (Cerac, purity = 99.5 %, 325 mesh), CNT (Nanocyl, purity> 95 %, size 10 nm - 1.5 µm) or a graphite modified powder (from Cerac one) and called in the following "graphene" according to the protocol describe by Paton *et al.* [43].

1.2. Mixing protocols

Three mixing methods have been used. Table 1 presents the synthesis conditions for each sample, in particular the mixing protocol. Sample No. 7 was synthesized for a previous test by Tusseau-Nenez *et al.* [14] and will serve as a reference throughout the experiment to check the reproducibility. Except for the "conventional protocol" (CP, described below), the uranium powder was dispersed in isopropanol under ultrasound then mixed with different carbon sources and with C/U ratios equal to 5, 6 or 7. In each case, 10 g of the final mixture were prepared.

The "conventional protocol", described previously [14], [39], consists in introducing into a 100 mL plastic bottle 5 grinding iron balls of 9 mm and the two precursor powders of uranium and carbon. The bottle is placed on an automatic mixer (Robin Power Mixer, PM031) for 12h.
The "developing protocol" (DP) is adapted from a protocol developed at ISOLDE [1], [23]. It consists in dispersing in an ultrasound bath and using a hand blender (Electrolux ESTM 955) a solution containing isopropanol and a CNT powder. Once the CNT have been dispersed, a solution of isopropanol containing uranium oxide (or oxalate) powder previously dispersed using ultrasound is introduced at the level of the agitation vortex. The agitation is maintained during 10 min in order to obtain a homogeneous mixture of the powders. Then the solution is placed in a water-bath with magnetic stirring and connected to a distillation column until total evaporation of the isopropanol. Once dried, this mixture of powders is manually ground in an agate mortar.

The "graphene protocol" (GP) aims to exfoliate graphite and is adapted from the method of Paton *et al.* [43]. In a blender (Magimix 11610), a water solution is mixed with 20 % sodium dodecyl sulphate (SDS) and graphite powder. During this mixing we added isopropanol to remove the foam formed. The previously dispersed uranium oxide powder is added to the blender. This mixture is then dried via a distillation column.

Table 1: Summary of the samples with the different qualitative variables used

| Samples | 12-day heating | C/U | Carbon source | Uranium source | Mixing protocol |
|---|---|---|---|---|---|
| No.1 $UO_2$ ground + CNT CP | no | 6 | CNT | $UO_2$ ground | Conventional Protocol |
| No.2 $UO_2$ ground + CNT DP | no | 6 | CNT | $UO_2$ ground | Developing Protocol |
| No.3 $UO_2$ ground + graphene GP | no | 6 | graphene | $UO_2$ ground | Graphene Protocol |
| No.4 OXA + graphite CP | no | 6 | graphite | OXA | Conventional Protocol |
| No.5 OXA ground + CNT DP | no | 6 | CNT | OXA ground | Developing Protocol |
| No.6 OXA + CNT DP | no | 6 | CNT | OXA | Developing Protocol |
| No.7 PARRNe BP894 | no | 6 | graphite | $UO_2$ ground | Conventional Protocol |
| No.8 PARRNe BP897 CP | no | 6 | graphite | $UO_2$ ground | Conventional Protocol |
| No.9 PARRNe BP897 CP 12d | yes | 6 | graphite | $UO_2$ ground | Conventional Protocol |
| No.10 $UO_2$ ground + CNT CP 12d | yes | 6 | CNT | $UO_2$ ground | Conventional Protocol |
| No.11 $UO_2$ ground + CNT DP 12d | yes | 6 | CNT | $UO_2$ ground | Developing Protocol |
| No.12 $UO_2$ ground + graphene GP 12d | yes | 6 | graphene | $UO_2$ ground | Graphene Protocol |
| No.13 $UO_2$ ground + CNT-5moles DP | no | 5 | CNT | $UO_2$ ground | Developing Protocol |
| No.14 $UO_2$ ground + CNT-7moles DP | no | 7 | CNT | $UO_2$ ground | Developing Protocol |

All the mixtures are pressed at 220 MPa for 6 sec in a 13 mm diameter mold with a semi-automatic press (SPECAC Automatic Hydraulic Press). The pellets were then carburized under vacuum ($10^{-6}$ mbar) at 1800 °C for 2 h. In order to check the stability of the obtained nanostructures at high temperature and over long period, i.e. corresponding to on-line experiments, samples No. 9/10/11/12 were further heat treated for 12 days at 1800 °C under vacuum.

2. Characterization techniques

All the samples were systematically characterized after carburization by X-Ray Diffraction (XRD - Bruker AXS, D8 Advance), helium pycnometry (Micromeritics, ACCUPYC 1330), mercury porosimetry (Micromeritics, Autopore IV 9500), Scanning Electron Microscopy (Zeiss, SEM-FEG Sigma HD) and Specific Surface Area measurement (SSA) by the Brunauer Emmett Teller method (BET - Micromeritics, ASAP 2020). The XRD analyses were performed as previously described [14]. The Rietveld refinement allowed us to determine the proportion of the UC and $UC_2$ phases as well as the crystallite sizes using the Maud program [44].

Helium pycnometry analysis was used to determine quantities of open and closed porosities and mercury porosimetry was used to determine the open pore size distribution. The details of the conditions for these two analytical techniques can be found in Hy *et al.* and Tusseau-Nenez *et al.* [14], [39]. The SEM observations combined with the SSA measurements allow determining the aggregate size and the size of the uranium carbide grains, respectively, without distinction between UC and $UC_2$.

SSA measurements are used to estimate the size of the grains. The SSA formula for a sample made of a mixture of compounds is written:

$$SSA = \frac{\sum n_i S_i}{\sum n_i m_i} \text{ (equation 1)}$$

with $n_i$, $S_i$ and $m_i$ the molar quantity, the developed surface and the mass of the compound i, respectively. The analyzed samples contain three phases: a carbon phase (graphite or CNT) and two uranium carbide phases (UC and $UC_2$). We consider the two uranium carbides identical in terms of morphology as they cannot be discriminated in SEM images. Therefore, we consider that they form only one phase called $UC_x$. So equation 1 can be written:

$$SSA = \frac{n_1.S_1 + n_2.S_2}{n_1.V_1.\rho_1 + n_2.V_2.\rho_2} \text{ (equation 2)}$$

with $n_{1\ or\ 2}$, $S_{1\ or\ 2}$, $V_{1\ or\ 2}$ and $\rho_{1\ or\ 2}$, the molar quantity, developed surface, volume and theoretical density of graphite or $UC_x$, respectively. Uranium carbide and graphite grains are supposed to be spherical. The SSA of the samples made from graphite is therefore written:

$$SSA = \frac{3.\left(r_1^2 + \frac{n_2}{n_1}.r_2^2\right)}{r_1^3.\rho_1 + \frac{n_2}{n_1}.r_2^3.\rho_2} \text{ (equation 3)}$$

with $r_{1\ or\ 2}$ the grain radius of graphite or $UC_x$, respectively.

The diameter of the graphite grains, 308 nm, was determined from SSA measurements made on this materials taking $\rho_{graphite}$ = 2.26 g.cm$^{-3}$. The theoretical density of the uranium carbide $UC_x$ ($\rho_2$ = 11.85 g.cm$^{-3}$) was calculated by weighting the theoretical densities of the UC ($\rho_{UC}$ = 13.65 g.mol$^{-1}$) and $UC_2$ ($\rho_{UC2}$ = 11.75 g.mol$^{-1}$) phases with the mass proportions determined by XRD (95 wt% of $UC_2$ and 5 wt% of UC, see Table 2)

For CNT the SSA was measured equal to 273 m$^2$.g$^{-1}$, which is in excellent agreement with the value given by the provider Nanocyl© (250-300 m$^2$.g$^{-1}$). The dimensions of the CNT given by the provider (diameter = 10 nm and length = 1.5 μm) were in consequence used in this study.
According to Peigney *et al.* [45], the SSA of multiwall CNT as follow:

$$SSA = \frac{\pi L d_c}{\frac{\pi L}{1315}\left(nd_c - 0.68 \sum_{i=1}^{n-1} i\right)} \text{ (equation 4)}$$

with $L$ the length of the CNT, $d_c$ its diameter and n the number of sheets.
We have deduced that the carbon nanotubes used contain each an average of 6 sheets. By modeling the $UC_x$ grains by spheres, the SSA of the samples prepared from CNT is written:

$$SSA = \frac{Ld_c + \frac{n_2}{n_1}4r_2^2}{\frac{L}{1315}(6d_c - 10.2) + \frac{4}{3}.\frac{n_2}{n_1}.r_2^3.\rho_2} \text{ (equation 5)}$$

with L the length of the CNT, $d_c$ its diameter, $r_2$ the radius of the $UC_x$ grain and $\rho_2$ the density of the $UC_x$. The grain sizes deduced from equations 3 and 5 are listed for each sample in table 2.

3. Characterization of materials and discussion

3.1. Grinding

As previously discussed in ref [40], the XRD analysis of the $UO_2$ powder after grinding shows a broadening and asymmetry of the diffraction lines compared to the raw powder (figure 1). This suggests the oxygen enrichment of the $UO_2$ phase leading to the formation of $UO_2$ phases enriched by oxygen which are more fragile. By grinding, an increase of the SSA from 3 $m^2.g^{-1}$ to 6.5 $m^2.g^{-1}$ is observed, corresponding to a decrease of the grain diameter (assumed to be spherical), from respectively 182 nm to 82 nm. These sizes are in agreement with SEM observations (figure 1b).

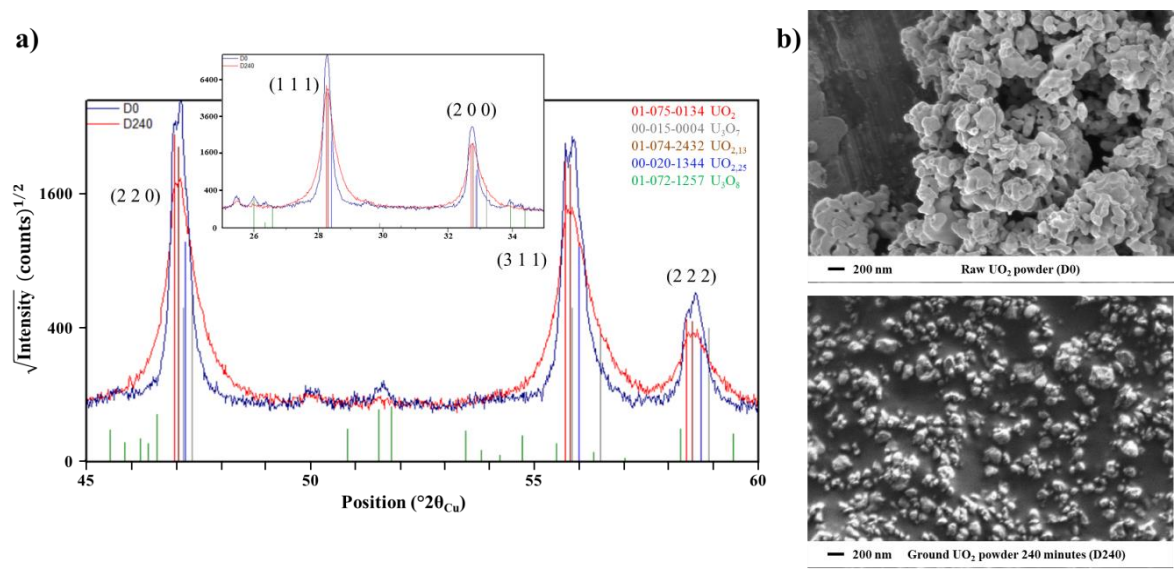

Figure 1: Influence of the grinding (240 min) on $UO_2$ powder a) XRD b) SEM images

The XRD analysis shows that uranium oxalate powder (OXA) is made of two hydrated phases: about 80% of $U(C_2O_4)_2,2H_2O$ and 20% of $U(C_2O_4)_2,6H_2O$ (figure 2). The ground OXA powder is partially dehydrated as only $U(C_2O_4)_2,2H_2O$ is identified by XRD. The corresponding Bragg peaks are much broader indicating a decrease in the crystallite size, to be linked to the grain size decrease observed by SEM (figure 2b). The very broad peak observed around 15 $°2\theta_{Cu}$ lets us to conclude that a partial amorphization occurs as related by Tyrpekl et al. even for temperatures below 100 °C [46].

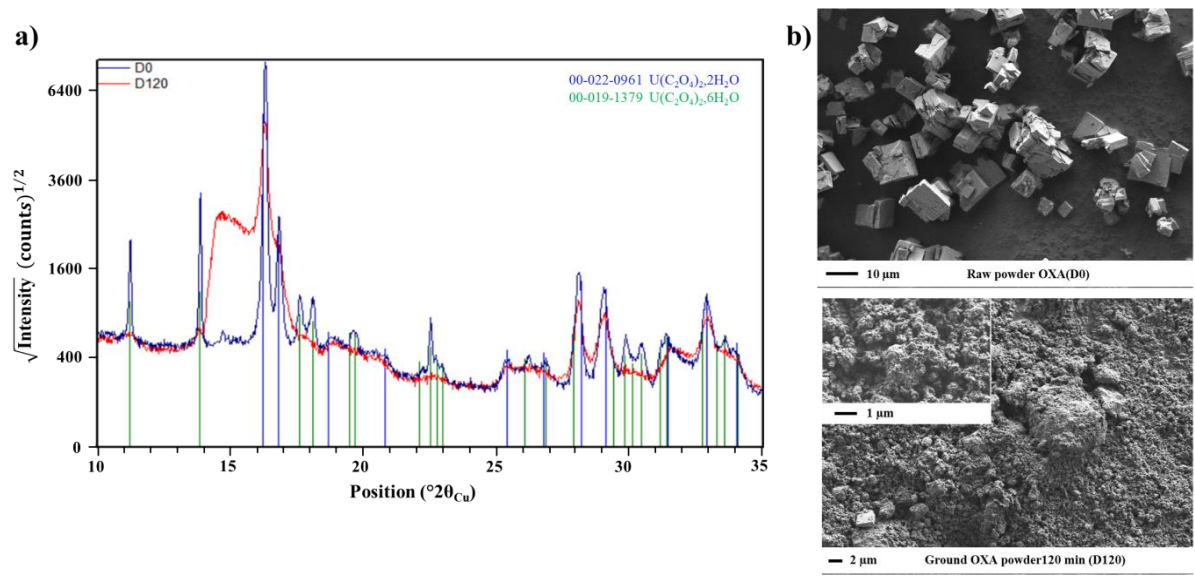

Figure 2: Influence of the grinding (120 min) on OXA powder a) XRD b) SEM images

3.2. Mixtures and carburization

Figure 3 shows a selection of SEM micrographs illustrating the influence of the mixing protocol. The results for the 14 samples are given in the complementary data. Clearly, the use of the conventional protocol induces the presence of aggregates of uranium oxide (or oxalate) and carbon in green pellets; that leads after carburization to the formation of large grains by abnormal growth. Therefore the interest of grinding is lost after sintering for this type of mixture. On the contrary, when the developing protocol is used, an intimate homogenization of uranium grains ($UO_2$ or OXA) in the matrix of CNT is obtained. This homogeneous microstructure is maintained after carburization and even after heating for 12 days. Finally, the "graphene" protocol allows inserting partially uranium grains between the graphite particles reduced in size and limits the appearance of aggregates. However, after heating for 12 days, the presence of large aggregates is locally observed.

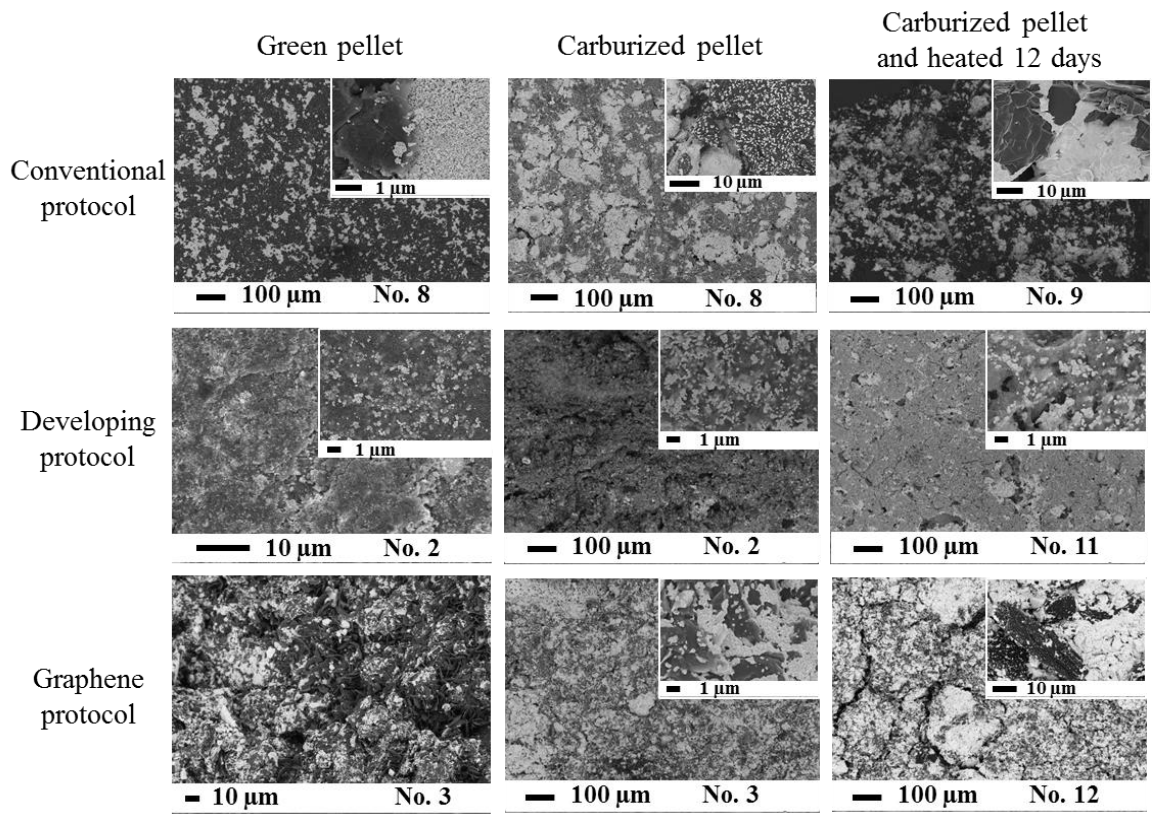

Figure 3: SEM images showing the influence of the mixing on the microstructure after carburization and 12 days isothermal heat treatment for three characteristic type samples.

Whatever the mixing protocol and the C/U ratio, the carburization leads to the formation of UC and $UC_2$ phases (figure 4). The graphite (or CNT) phase is identified as it was added in excess. The relative weight proportions of uranium carbides are calculated to be 95 wt% of $UC_2$ and 5 wt% of UC. In addition, the crystallite sizes are $52 \pm 16$ nm and $127 \pm 24$ nm respectively for UC and $UC_2$. These values are the averages and the standard deviations obtained from the measurements performed on the 14 samples. In comparison with SEM observation, these results mean that the two uranium carbide phases have polycrystalline grains.

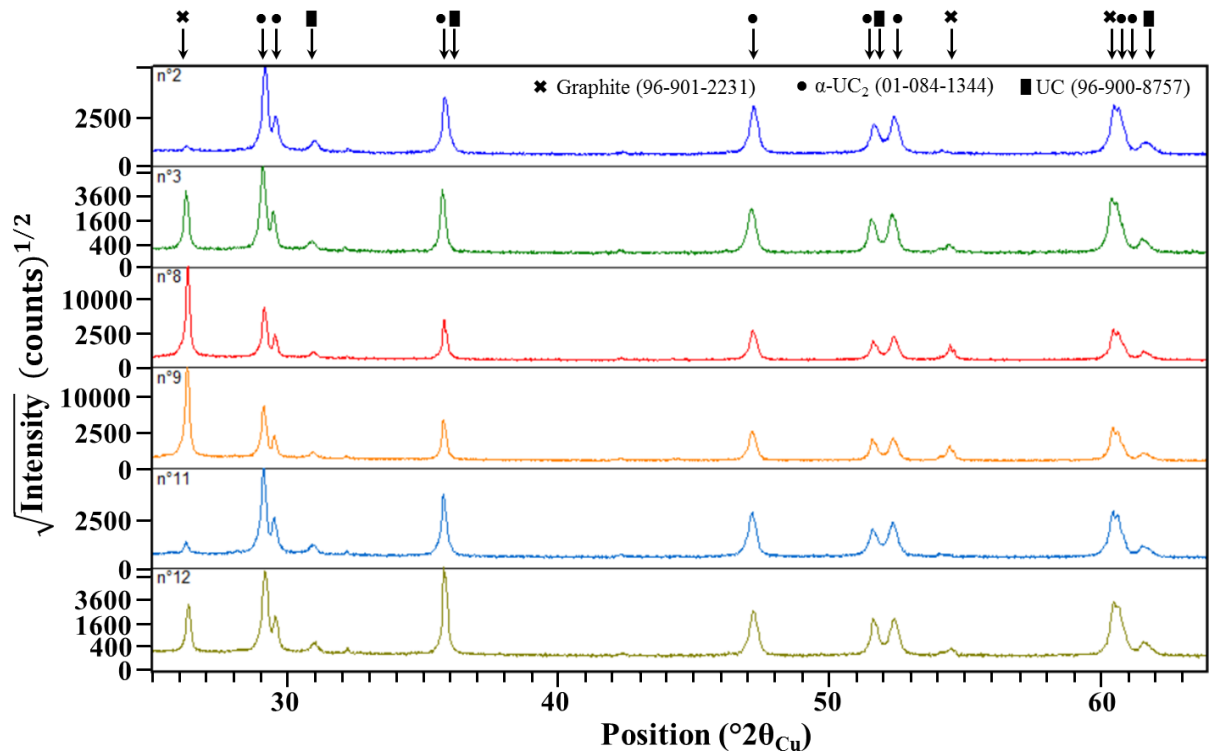

Figure 4: XRD phase identification for three characteristic type samples after carburization and with or without a 12-days isothermal heat treatment.

In conclusion, by using all these characterization techniques, the structure and the microstructure of our 14 carburized samples can be described by 14 physicochemical variables presented in table 2.

The results obtained previously [14], [39] indicated links between the synthesis conditions (nature of the precursors, grinding, carburization) and the structure and microstructure of the sintered pellets. In our present study we aim to establish real correlations and in order to minimize subjective bias, we have performed a multivariate statistical analysis.

4. Statistical analysis by Principal Component Analysis

The Principal Component Analysis (PCA) is a multivariate statistical method that allows us to analyze the information contained in a data set, to identify the linear correlations existing between the variables and to synthesize this information by reducing the variable number [47], [48]. In order to study the impact of variables expressed in different units, the correlation matrix has to be computed using the standardized (i.e. centered and scaled) data. Diagonalizing this matrix gives an access to the principal components that are the unit eigenvectors of the correlation matrix. The PCA provides a graphical representation of the data set in a two-dimensional space defined by the first two principal components (called Dim1 and Dim 2 in the following graphs).

The PCA was carried out using the 14 physicochemical variables presented in Table 2 as quantitative variables and the mixing protocols, the nature of the uranium and carbon sources, the C/U ratio and the heating after carburization as qualitative variables. Table 2 shows that the variables related to the phase nature and their proportions ($UC_2$, UC and C) have a small standard deviation, this originates from the fact that most of the samples were synthesized with the same initial molar ratio C/U = 6, except the No. 13 and 14 samples. The standard deviation measures the dispersion of a variable. So, when the standard deviation of a variable involved in the sample description is small, the samples show little variability for this variable, in other words, this variable provides little information to discriminate between the samples. Therefore, the %$UC_2$, %UC and %C variables were considered as supplementary variables, this means they do not contribute to the construction of the principal components, contrary to the active variables. As for qualitative variables, they must be treated as supplementary in a PCA.

Table 2: Summary of the carburized samples with the different quantitative variables used. $P_{35}$, $P_{200}$, $P_{3000}$, $P_{10000}$ and $P_{30000}$ represent the percentages of open porosity on the diameter pores 0.035 μm, 0.2 μm, 3 μm, 10 μm and 30 μm, respectively.

|  | XRD* |  |  |  |  | BET | SEM | He Pycnometry |  | Hg Porosimetry |  |  |  |  |
|---|---|---|---|---|---|---|---|---|---|---|---|---|---|---|
|  | Phase and proportion (%, ± 1 %) |  |  | Crystallite size (nm, ± 5 nm) |  | UCx Grain size (nm)** | UCx Aggregate size (μm) | Porosity (%, ± 1 %) |  | Open pore size Distribution (%) |  |  |  |  |
|  | UC | $UC_2$ | C | UC | $UC_2$ |  |  | Open | Close | $P_{35}$ | $P_{200}$ | $P_{3000}$ | $P_{10000}$ | $P_{30000}$ |
| No.1 $UO_2$ ground + CNT CP | 3 | 88 | 9 | 59 | 87 | 118 | 15 | 78 | 7 | 22 | 10 | 10 | 0 | 58 |
| No.2 $UO_2$ ground + CNT DP | 5 | 86 | 9 | 39 | 114 | 100 | 0.5 | 68 | 12 | 34 | 32 | 34 | 0 | 0 |
| No.3 $UO_2$ ground + graphene GP | 4 | 88 | 8 | 51 | 129 | 1200 | 18 | 49 | 7 | 3 | 22 | 56 | 0 | 19 |
| No.4 OXA + graphite CP | 4 | 87 | 9 | 55 | 160 | 820 | 23 | 55 | 5 | 2 | 12 | 86 | 0 | 0 |
| No.5 OXA ground + CNT DP | 13 | 78 | 9 | 40 | 127 | 94 | 0.6 | 70 | 15 | 24 | 29 | 47 | 0 | 0 |
| No.6 OXA + CNT DP | 7 | 84 | 9 | 65 | 149 | 82 | 3.2 | 74 | 14 | 17 | 17 | 66 | 0 | 0 |
| No.7 PARRNe BP894 | 5 | 86 | 9 | 102 | 165 | 906 | 31 | 41 | 5 | 4 | 14 | 82 | 0 | 0 |
| No.8 PARRNe BP897 CP | 5 | 87 | 8 | 38 | 145 | 972 | 65 | 51 | 5 | 2 | 8 | 46 | 44 | 0 |
| No.9 PARRNe BP897 CP 12d | 5 | 87 | 8 | 46 | 144 | 914 | 56 | 49 | 8 | 2 | 5 | 49 | 44 | 0 |
| No.10 $UO_2$ ground + CNT CP 12d | 3 | 88 | 9 | 48 | 86 | 100 | 71 | 72 | 13 | 19 | 13 | 10 | 0 | 58 |
| No.11 $UO_2$ ground + CNT DP 12d | 5 | 86 | 9 | 42 | 110 | 96 | 0.5 | 64 | 17 | 30 | 33 | 37 | 0 | 0 |
| No.12 $UO_2$ ground + graphene GP 12d | 4 | 88 | 8 | 43 | 135 | 1412 | 14 | 48 | 4 | 3 | 17 | 65 | 0 | 15 |
| No.13 $UO_2$ ground + CNT-5moles DP | 5 | 90 | 5 | 57 | 119 | 104 | 0.6 | 64 | 8 | 29 | 27 | 35 | 0 | 9 |
| No.14 $UO_2$ ground + CNT-7moles DP | 4 | 84 | 12 | 40 | 102 | 92 | 0.2 | 69 | 15 | 33 | 37 | 30 | 0 | 0 |
| Standard deviation | 2 | 3 | 1 | 16 | 24 | 484 | 26 | 11 | 4 | 12 | 10 | 22 | 15 | 20 |

* For all the samples, the agreement factors were in the ranges: 11.6% < $R_w$ < 14.8%, 5.8% < $R_{exp}$ < 7.3%, 1.9 < $\chi^2$ < 2.3.
** Error bar of SSA measurements is 5%

The PCA results are presented in figure 5: a) the graph of variables and b) the graph of samples. It appears that the plane defined by the first two principal components explain 72.42 % of the information contained in the data set. This figure is higher than the reference value that equals 50.80 %, the reference value being the 0.95 quantile of the inertia percentage distribution obtained by simulating 10000 data tables of equivalent size assuming independent variables [48].

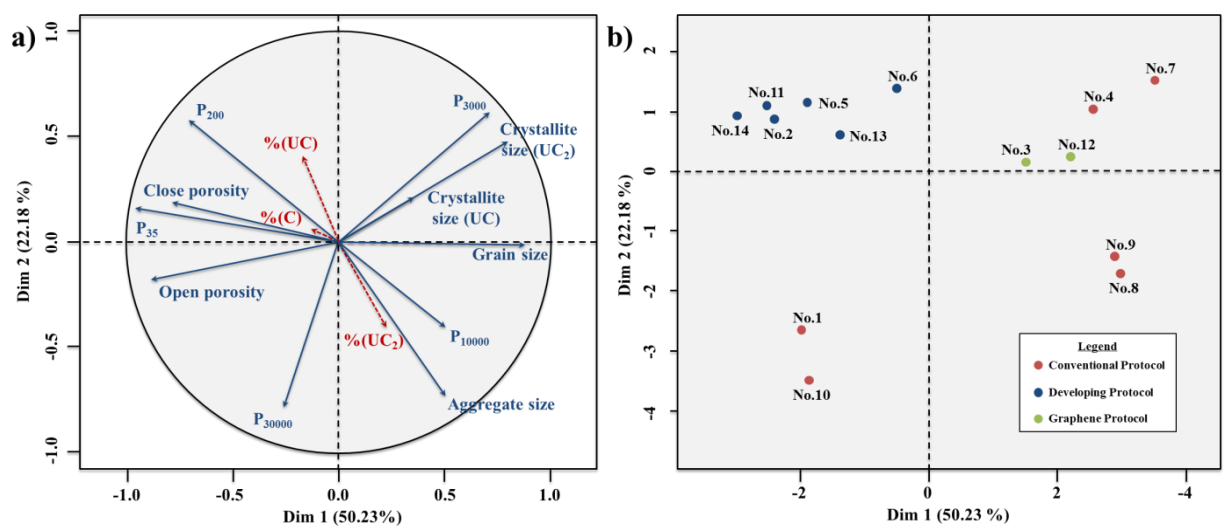

Figure 5: PCA results a) variable graph (only blue variables were taken into account) b) Sample graph

The graph of variables, also called correlation circle, shows the correlation between the variables and the first two principal components. The length of the vector representing a variable measures the quality of the representation (called $\cos^2$) of this variable in the principal plane. Variables with a vector arrow close to the edge of the circle, i.e. with a $\cos^2$ value close to 1, are well represented in the principal plane. In figure 5a, this is the case for the active variables exhibiting $\cos^2$ values greater than 0.7: all except the two ones called "crystallite size UC" and "$P_{10000}$". The first principal component, called Dim. 1, is positively correlated with the variables "Grain size", "Crystallite sizes $UC_2$" and the open pore size "$P_{3000}$", and negatively correlated with the variables "open and closed porosity" and small pores "$P_{200}$" and "$P_{35}$". The variables that best describe the second principal component (Dim. 2) are the pore size variables "$P_{3000}$" and "$P_{200}$" and the variables "Aggregate size" and "$P_{30000}$". The correlation with the first two variables is positive and negative with the last two ones.

The first two principal components explained in terms of the active variables can be used to describe the graph of samples also called graph of individuals (figure 5b):

- In the top left, are expected the samples with a small component along Dim.1 and a large component along Dim.2, that means with a structure characterized by small grains, a high porosity distributed over pores with small diameters (35 and 200 nm) and small aggregates. These features correspond to the samples No. 2, 5, 6, 11, 13 and 14.
- In the bottom left, are displayed the samples with small components along both dimensions. These samples have small grains forming large aggregates, a high porosity distributed mainly over pores with large diameters (30 µm). These characteristics correspond to the samples No. 1 and 10.
- In the top right, the samples have large components along the two dimensions. This implies large grains, a small porosity mainly spread over pores with a 3 µm diameter. These attributes correspond to samples No. 3, 4, 7 and 12.
- In the bottom right, the samples show a large component along the first axis and a small one along the second axis. These samples are characterized by large grains and aggregates and a small porosity mainly distributed over pores with 3 and 10 µm diameters, namely the samples No. 8 and 9.

It is worth noting that, except samples No. 3, 5 and 12, all individuals are well represented in the principal plane since they have a $\cos^2 > 0.6$. The samples No. 7 and 8 were expected to have very similar characteristics since sample No. 7 is the reference one and sample No. 8 was synthesized following the same protocol. They show very similar grain sizes but different aggregate sizes, highlighting the limits of reproducibility of the conventional protocol (CP).

The PCA also provides information on qualitative variables considered as supplementary variables. Figure 6 shows, in the principal plane, the confidence ellipses drawn around the modalities of each qualitative variable, i.e. around the barycenter of the samples having this modality. For example, for the qualitative variable "Carbon source", the three modalities are "graphite", ''graphene'' and "CNT". The green square labelled ''graphite'' is the barycenter of the samples No. 4, 7, 8 and 9, the pink square noted ''graphene'' is the barycenter of the samples No. 3 and 12 and the blue square labelled ''CNT'' the barycenter of the samples No. 1, 2, 5, 6, 10, 11, 13 and 14. In the case of graphene, the ellipse is reduced to a line since only 2 samples define the modality. The 3 modalities "graphite", "graphene" and "CNT" are well separated on the graph. However, graphite and graphene sources are quite close to each other, which reflects the structural proximity of these two carbon sources. The pattern obtained for the "Carbon source" variable points out that all the samples synthesized with CNT are located in the left side of the individual graph, i.e. have a negative component on Dim.1, contrary to the samples based on graphite or graphene that have a positive component along Dim.1 and thus appear in the right side of the graph.

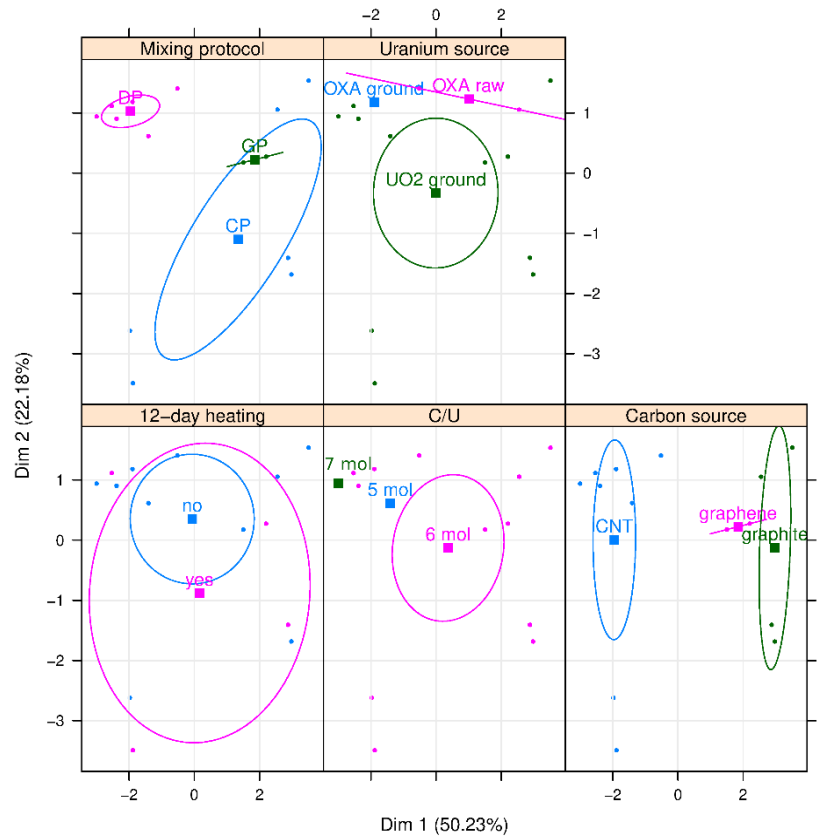

Figure 6: PCA results with qualitative variables

For the "uranium source" variable, the modalities "OXA raw" and "OXA ground" are not well-differentiated, but the modality "$UO_2$ ground" is well separated from the two others.

Concerning the C/U variable, that is the molar ratio between the carbon and uranium quantities, the modalities "C/U" equal to 5 and 7 moles concern only one sample (No. 13 and 14, respectively). They appear in figure 6 by only a square. They are well represented in the plane (1, 2) with a cos² value equal to 0.89 and 0.60 respectively. These two modalities are significantly separated from each other and from the ellipse corresponding to C/U = 6, indicating a statistically significant influence of the C/U variable.

As for the ''12-day heating'' qualitative variable, the confidence ellipse of the modality "no" is fully included in the "yes" ellipse, indicating that the two modalities are not significantly different. This is in agreement with the graph of individuals where the coordinates of samples synthesized according to the same protocol, but with or without isothermal heating at high temperature after carburization (for example No. 2 and 11, No. 1 and 10 or No. 8 and 9) are very close (see figure 5b). This demonstrates that the microstructure is not modified by a 12-day thermal treatment.

Finally, for the qualitative variable "Mixing protocol", the DP modality is well separated from the two others. This corroborates the influence of the mixing protocol already evident in figure 5b where all the samples prepared according to the DP protocol are located in the top left-hand side. This variable therefore plays a significant role in the microstructural properties.

In conclusion, the PCA has demonstrated that using CNT is the key parameter to synthesize samples with small grains and high porosity and that the DP mixing protocol is necessary to limit the aggregate size and to obtain an open porosity mainly distributed over pores with small diameters (35 and 200 nm). The microstructure obtained appears to be stable at high temperature since it is not modified when the samples are held at 1800 °C for 12 days. This seems to indicate that the grain growth occurred during the carburization.

5. Conclusion

In this study, 14 different samples were prepared with different precursor powders ($UO_2$ or uranium oxalate on the one hand and graphite, graphene or CNT on the other hand) and different mixing protocols (CP, DP and GP). Some samples underwent an additional long-time heating treatment after carburisation in order to test the high-temperature stability of the obtained microstructures. The samples were systematically analysed using various characterisation technics in order to determine the proportions of the phases obtained after carburisation and the size of the crystallites (XRD), the size of the grains and aggregates (BET and SEM), the proportions of the open and closed porosities (He pycnometry) and the open-pore size-distribution (Hg porosimetry). Each sample is described by 19 variables: 5 qualitative variables related to the synthesis process and 14 quantitative variables connected with its physicochemical properties. The results have been interpreted in the frame of a multivariate statistical approach using the principal component analysis. This statistical procedure removes any subjective bias in the data analysis and allows a graphical representation of the correlations between the variables.

The PCA demonstrated that using the CP protocol leads systematically to a heterogeneous microstructure characterized by different aggregate sizes; on the contrary, when the DP protocol is applied, the targets synthesized from milled $UO_2$ and CNT have a homogeneous microstructure stable at high temperature.

The PCA has also revealed the strong correlations between the sample-preparation methods, the dwell time after carburisation and the structural properties, in particular the role of the porosity (quantity and distribution on small pores), and of the grain and aggregate sizes is firmly established, consolidating the results of the characterisations. Using CNT and applying the DP protocol leads to limit the sintering and obtain samples with small grains and a high porosity spread over open pores with small diameters (35 and 200 nm).

This study is the first step to establish a correlation between the physicochemical properties of the targets and their FPs release behaviour.

**Acknowledgments:**

The authors would like to thank Ulli Köster, Pascal Jardin and Olivier Tougait for the precious and fruitful discussions. We thank the ALTO team for helping us to install the implementation devices. We also thank Kevin Dolin-Dolcy for his involvement in the project during his internship in 2015.

**Complementary data:**

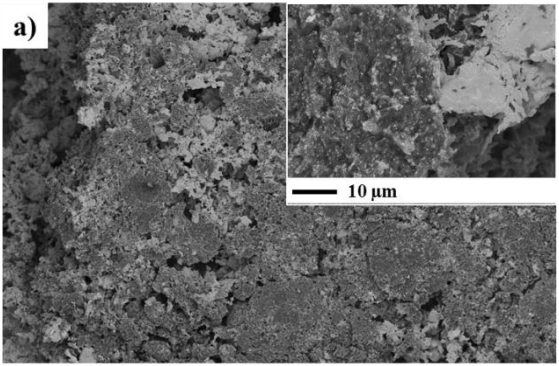
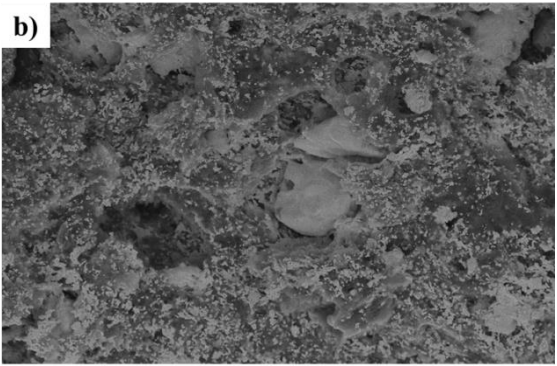
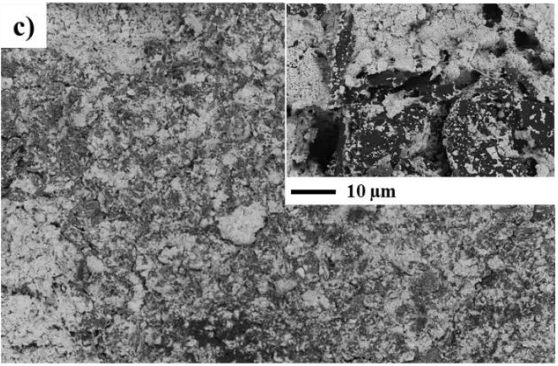
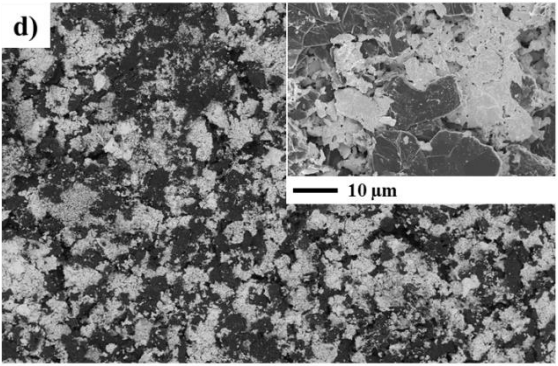
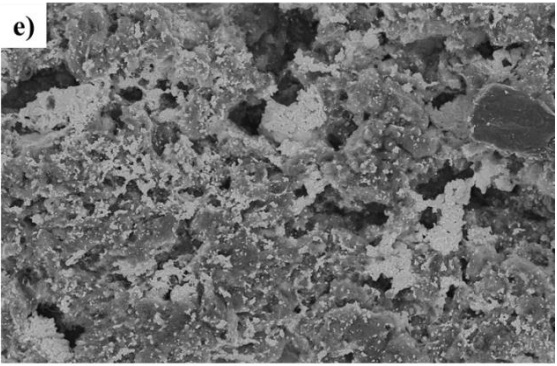
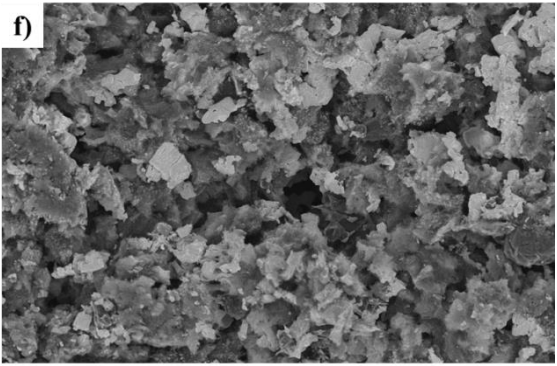
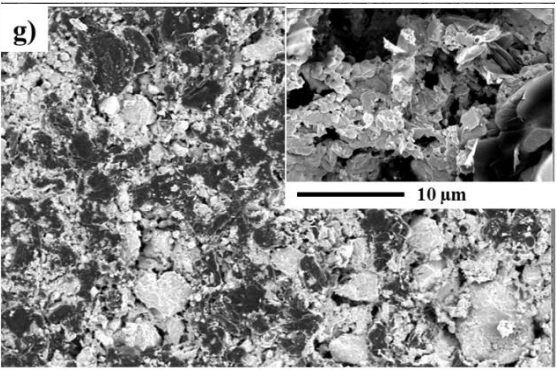
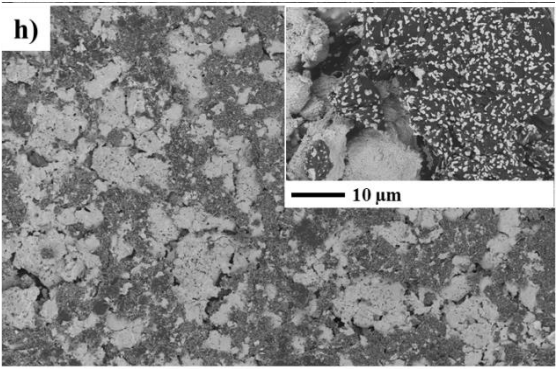

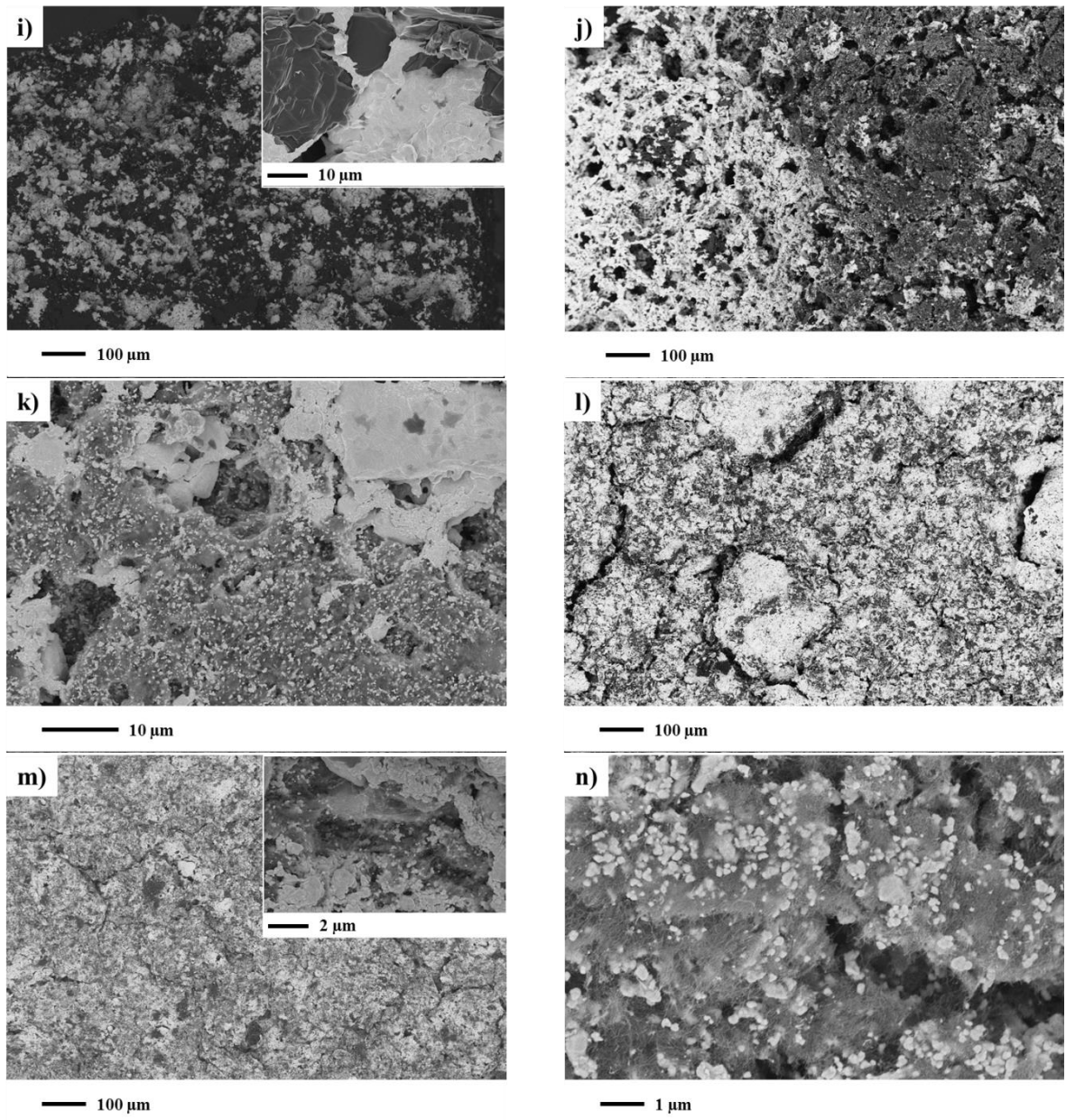

Figure 66: SEM images of the 14 carburized samples

- a) No.1 UO$_2$ ground + CNT CP
- b) No.2 UO$_2$ ground + CNT DP
- c) No.3 UO$_2$ ground + graphene GP
- d) No.4 OXA + graphite CP
- e) No.5 OXA ground + CNT DP
- f) No.6 OXA + CNT DP
- g) No.7 PARRNe BP894
- h) No.8 PARRNe BP897 CP
- i) No.9 PARRNe BP897 CP 12d
- j) No.10 UO$_2$ ground + CNT CP 12d
- k) No.11 UO$_2$ ground + CNT DP 12d
- l) No.12 UO$_2$ ground + graphene GP 12d
- m) No.13 UO$_2$ ground + CNT-5moles DP
- n) No.14 UO$_2$ ground + CNT-7moles DP